\documentclass[graybox, secnum]{svmult}

\usepackage{mathptmx}       % selects Times Roman as basic font
\usepackage{helvet}         % selects Helvetica as sans-serif font
\usepackage{courier}        % selects Courier as typewriter font
\usepackage{type1cm}        % activate if the above 3 fonts are
\usepackage{makeidx}         % allows index generation
\usepackage{graphicx}        % standard LaTeX graphics tool
\usepackage{multicol}        % used for the two-column index
\usepackage[bottom]{footmisc}% places footnotes at page bottom
\usepackage{hyperref}        %for hyperlinks
\usepackage{soul}            % for high-lighting of text
\hypersetup{colorlinks=true,urlcolor=blue}

\usepackage{float}

\usepackage[square,numbers]{natbib}
\makeindex             % used for the subject index
\begin{document}
%\tableofcontents{}
\title*{X-ray Optics for Astrophysics: a historical review}
\author{Finn E. Christensen and Brian D. Ramsey\thanks{corresponding author}}
\institute{Finn E. Christensen \at DTU Space, Technical University of Denmark, Elektrovej, Bygn. 327, Lyngby, Denmark , \email{finn@space.dtu.dk}, 
\and Brian D. Ramsey \at NASA Marshall Space Flight Center, Huntsville, AL 35812 (United States), \email{Brian.Ramsey@nasa.gov}}
\maketitle
\abstract{ Grazing-incidence X-ray optics have revolutionized X-ray astrophysics. The ability to concentrate flux to a tiny detection region provides a dramatic reduction in background and a consequent very large improvement in sensitivity. The X-ray optics also permit use of small-format, high-performance focal plane detectors and, of course, especially for high-angular-resolution optics, provide a wealth of imaging data from extended sources. This review, follows the use of X-ray optics from the first rocket-borne instruments in the 1960s through to the Observatories flying today and being developed for future use. It also includes a brief overview of the challenges of fabricating X-ray optics and the various technologies that have been used to date.}
\section{Keywords} 
Grazing-incidence optics, X-ray telescope, X-ray imaging, X-ray Astronomy.

\section{Early Days of X-Ray Astronomy}
%\begin{document}
The development of rockets and later satellites made possible the observation of celestial X-rays from outside the earth’s attenuating atmosphere. However, although X-ray emission had been detected from the sun as early as the 1940s\cite{burnight1949soft}  it was expected that if nearby stars produced similar amounts of X-rays their signal would be $\sim 10$ orders of magnitude smaller, way beyond the capabilities of instrumentation at that time, and therefore no extra-solar X-ray sources could be expected to be detected. All this changed during a very brief rocket flight nearly 60 years ago\cite{giacconi1962evidence}. Intended to look for X-ray fluorescence from the moon, this flight discovered an extremely bright source of X rays from the constellation Scorpius as well as a diffuse isotropic X-ray background and marked the beginning of the field of X-ray astronomy. It soon became apparent through other sub-orbital measurements\cite{bowyer1965cosmic} that there were many powerful X-ray sources (10 orders of magnitude brighter in X rays than the sun), that some were extended, and that they were nearly all time variable on a variety of scales. These discoveries were made with simple instruments with mechanical collimators to provide relatively crude locations on the sky. 

\begin{figure} [ht]
%   \begin{center}
\centering
   \begin{tabular}{l} %% tabular useful for creating an array of images 
   \graphicspath{ {./images/} }
   \includegraphics[width=8.0cm]{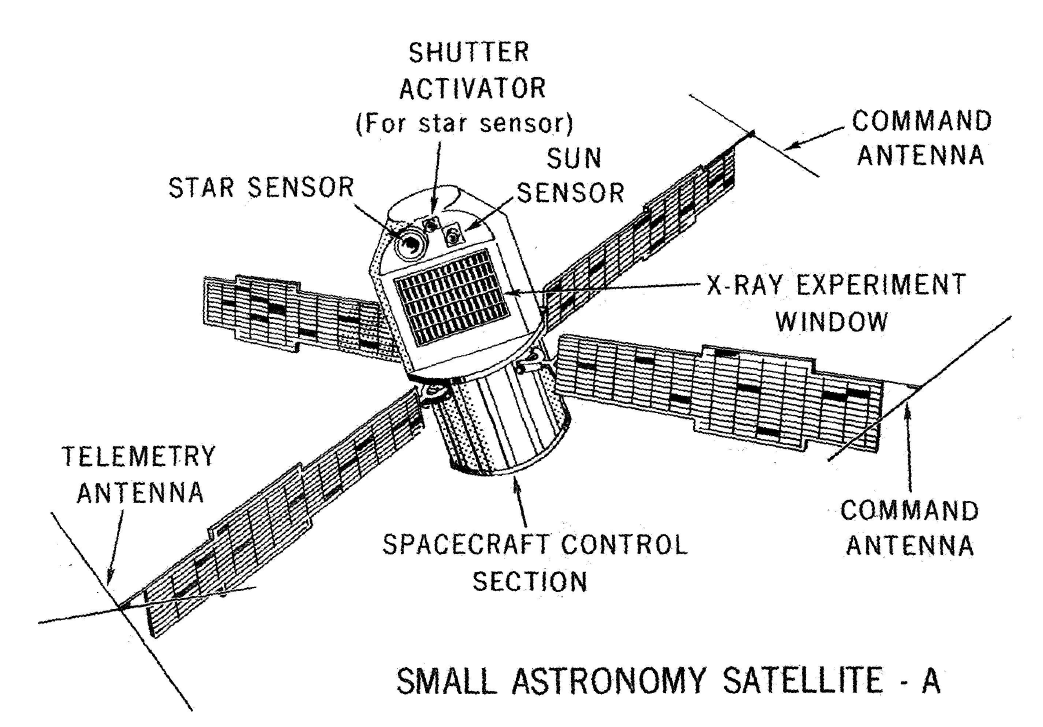}
   \end{tabular}
%   \end{center}
   \caption[Uhuru] 
%>>>> use \label inside caption to get Fig. number with \ref{}
   { \label{fig:Uhuru} The UHURU Satellite (SAS-A)}
   \end{figure}

\begin{figure} [ht]
   \begin{center}
   \begin{tabular}{l} %% tabular useful for creating an array of images 
   \graphicspath{ {./images/} }
   \includegraphics[width=9.0cm]{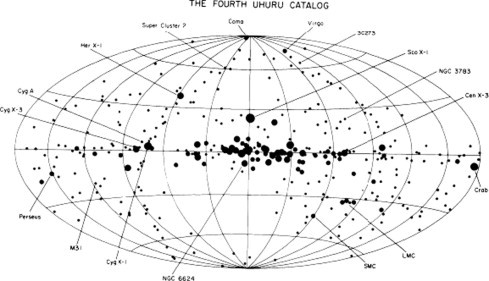}
   \end{tabular}
   \end{center}
 
  \caption[uhuruMap] 
   %>>>> use \label inside caption to get Fig. number with \ref{}
  % { \label{fig:uhuruMap} The UHURU x-ray source catalog}
   {\label{fig:uhuruMap} The UHURU X-ray source catalog}
   \end{figure}

In 1970, the first satellite dedicated to X-ray astronomy began operation. Called UHURU\cite{giacconi1971x}, the Swahili word or freedom, it was launched from Kenya into an equatorial orbit and operated for just over 2 years. UHURU (Figure \ref{fig:Uhuru}) used mechanically-collimated gas-filled detectors with a narrow field of view and rotated slowly (0.08 rpm) so that a sources position in the scan direction could be determined from the satellite’s pointing information. Repeating this process in several scan directions leads to a source’s location on the sky. The angular resolution obtained by UHURU depended on the brightness of the source and also the local density of other sources which could confuse the observation. For isolated bright sources UHURU achieved localization accuracies of less than 1 arcminute. Over its 27 months of operation, UHURU produced a catalogue of 339 X-ray sources (Figure \ref{fig:uhuruMap}).

%\begin{figure} [p]
%\centering
%\begin{subfigure}[c]{.5\linewidth}
%\centering
%\includegraphics[width=\linewidth]{uhuru.png}
%\end{subfigure}
%\hfill
%\begin{subfigure}[c]{.5\linewidth}
%\centering
%\includegraphics[width=\linewidth]{uhuru_map.jpg}
%\end{subfigure}
%\caption{\label{fig:uhuru} UHURU satellite (top) and X-ray source catalogue (bottom)}
%\end{figure} 

The success of UHURU fueled the desire for ever more sensitive measurements and other experimental techniques were explored. Large-area detectors are limited in sensitivity by their inherently large backgrounds, which can approach the signals of the brightest cosmic X-ray sources. Only by performing on/off source measurements for extended periods of time can faint sources be detected. The potential for using X-ray telescopes that could focus source flux to a small detector area had already been proposed even before the discovery of the first cosmic X-ray source\cite{giacconi1960telescope}. As we shall see, it was this development that was to revolutionize the field of X-ray astronomy.
%\end{document}

\section{The Benefit of X-Ray Optics}
%\begin{document}
\subsection{Signal to Noise Advantage}
Detector background consists of multiple components. Cosmic rays, especially trapped particles in the earth’s radiation belts an interact directly with the X-ray detector, or indirectly by producing secondaries, typically gamma rays, that then interact in the detector sensitive volume. In addition, there is a diffuse isotropic cosmic X-ray background that depends on the field of view of the instrument. All these components are dependent on the detector volume.

The sensitivity of a detector is given by $Sens = S / \sqrt{S + B}$ where $S$ is the total source counts and $B$ is the background counts.  As the background typically dominates this reduces to $S / \sqrt B$ and as both source and background counts increase with detector area the sensitivity scales as $\sqrt A$. Thus to double the sensitivity of a detector its area would have to be 4 x larger and this quickly leads to a limiting practical sensitivity.

If an X-ray mirror is used, then the X-ray flux is focused on to a small detection area, thereby greatly reducing the background and dramatically improving the signal to noise. The source and background flux are essentially decoupled as the source flux can be increased by increasing the mirror collecting area (for a fixed angular resolution and focal length). The background can be decreased by reducing the focal spot size on the detector. By way of an example, consider the Chandra observatory, now flying (see \S\ref{1990s}). The effective area of Chandra is approximately 800 cm$^2$ at 0.25 keV, approximately comparable to that of the UHURU observatory (each of 2 sets of proportional counters had 840 cm$^2$ effective area). However, the optics, which have an angular resolution of less than 0.5 arcsec HPD on axis\cite{POG}, focus the flux down to just a  $\sim 25$-micron-square area, giving rise to a background rate of just a few counts per year\cite{chandraACIScalibration}, compared to typical large area detector rates of a few counts per second. The net result is that Chandra has a sensitivity a factor of 10$^5$ greater than UHURU, and in its deep field images has detected $\sim1000$ sources in a 0.13 deg$^2$ region of the sky\cite{luo2016chandra}, compared with 339 sources in the whole sky with UHURU. 

\subsection{Large Dynamic Range and Less Source Confusion}
Because the signal from a source is concentrated in one area of the detector when using focusing optics this leads to the capability to observe faint sources (or source structure) in the presence of bright ones This is in sharp contract to a non-imaging system where each visible source contributes across the full detector area, and thus faint sources become lost in the noise of brighter sources and the background.
\subsection{Use of High Performance Detectors}
X-ray optics permits the use of small-format, high-performance focal-plane detectors that represent the state of art developments. Early large-area detectors were typically gas-filled proportional counters that had modest energy resolution (1.5 keV at 5.9 keV). The use of mirror systems allowed CCDs to become ubiquitous in X-ray astronomy, with an almost factor of 10 improvement in both energy resolution and spatial resolution. When coupled with diffraction gratings placed in the optical path, spectral resolutions of the order of 1 eV are now possible at low energies. Recent advances in focal-plane detectors include CMOS direct imagers in which each imaging pixel has its own electronics read-out chain for much faster readout rates, and microcalorimeters which offer the promise of high spatial and spectral resolution in the 0.2-7
keV energy range\cite{bandler2019lynx}.
%\end{document}

\section{The Challenges of Fabricating X-ray Optics}

%\begin{document}
\subsection{X-ray reflection}
\begin{figure} [ht]
   \begin{center}
   \begin{tabular}{l} %% tabular useful for creating an array of images 
   \graphicspath{ {./images/} }
   \includegraphics[width=9.0cm]{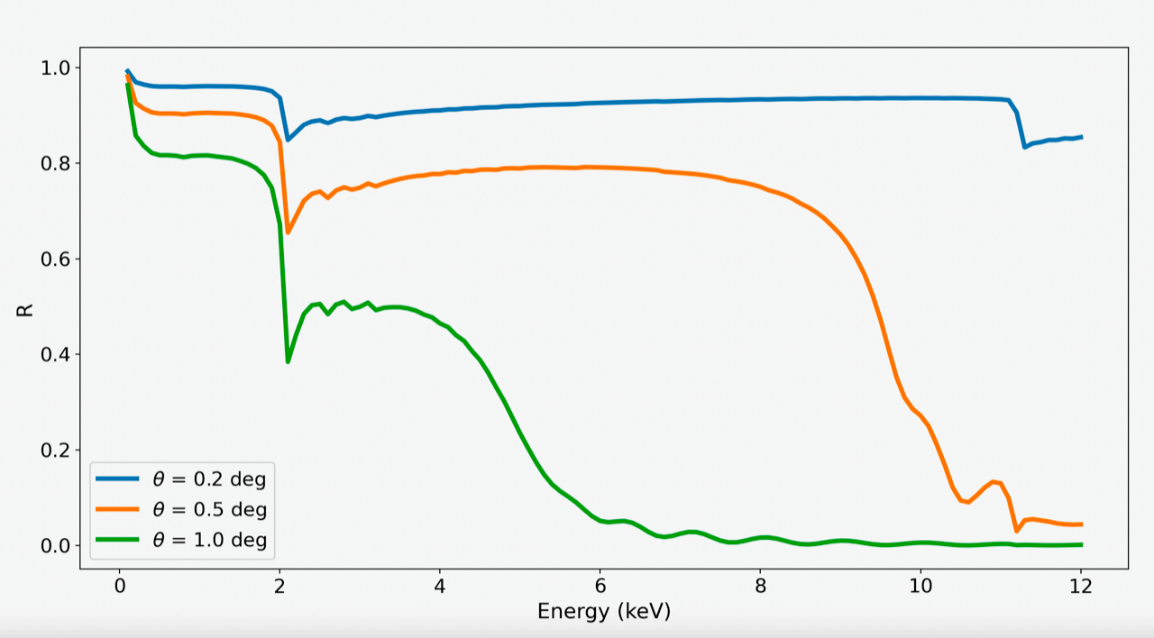}
   \end{tabular}
   \end{center}
   \caption[iridiumReflectivity] 
%>>>> use \label inside caption to get Fig. number with \ref{}
   { \label{fig:iridiumReflectivity} Reflectivity of iridium as a function of energy for 3 graze angles}
   \end{figure}

For X rays, the refractive index of metals is just below one and thus total external reflection can take place at the surface of the material. The critical angle $\theta_c$ , below which reflection occurs, can be calculated from dispersion theory and away from absorption edges is approximately given by:
\begin{equation}
    \theta_r = \left( 4 \pi r_0 \lambda^2 n\right)^{0.5}
\end{equation}
where $r_0$ is the classical electron radius, $n$ is the electron density and $\lambda$ the X-ray wavelength\cite{aschenbach1985x}. The largest so-called ‘graze’ (for grazing incidence) angles are therefore at the largest wavelength and with the most dense materials. Typical values are 1$^{\circ}$ or less for iridium or gold over the 1-10 keV energy range. These high-Z reflective materials would normally be coated on the mirror surface, via vacuum deposition, to enhance reflectivity. A typical reflectivity curve, for iridium, is shown in Figure \ref{fig:iridiumReflectivity}.

\subsection{Optical Configuration}
\begin{figure} [ht]
   \begin{center}
   \begin{tabular}{l} %% tabular useful for creating an array of images 
   \graphicspath{ {./images/} }
   \includegraphics[width=9.0cm]{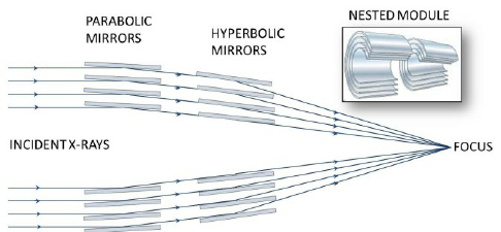}
   \end{tabular}
   \end{center}
   \caption[wolter1] 
%>>>> use \label inside caption to get Fig. number with \ref{}
   { \label{fig:wolter1} Wolter-1 nested optical configuration}
   \end{figure}

The phenomenon of total external reflection can be used to make an X-ray mirror. Parabolic reflectors with cylindrical cross sections and shallow (interior) graze angles were initially proposed\cite{giacconi1960telescope} which could provide very good on axis angular resolution, but would have off-axis responses strongly affected by coma. Later use was made of combinations of surface shapes and the so-called Wolter-1 geometry\cite{wolter1952spiegelsysteme}, which combines a parabolic first segment followed by a hyperbolic second segment (typically of equal lengths), has become widely used in X-ray astronomy\cite{vanspeybroeck1972design}. Because projected mirror collecting areas are quite small, due to the shallow graze angles, and because the mirror shells cannot be increased to arbitrary lengths without significantly degrading angular resolution,  multiple mirror shells with the same focal length are concentrically nested to increase effective area, as shown in Figure \ref{fig:wolter1}.

\begin{figure} [ht]
   \begin{center}
   \begin{tabular}{l} %% tabular useful for creating an array of images 
   \graphicspath{ {./images/} }
   \includegraphics[width=9.0cm]{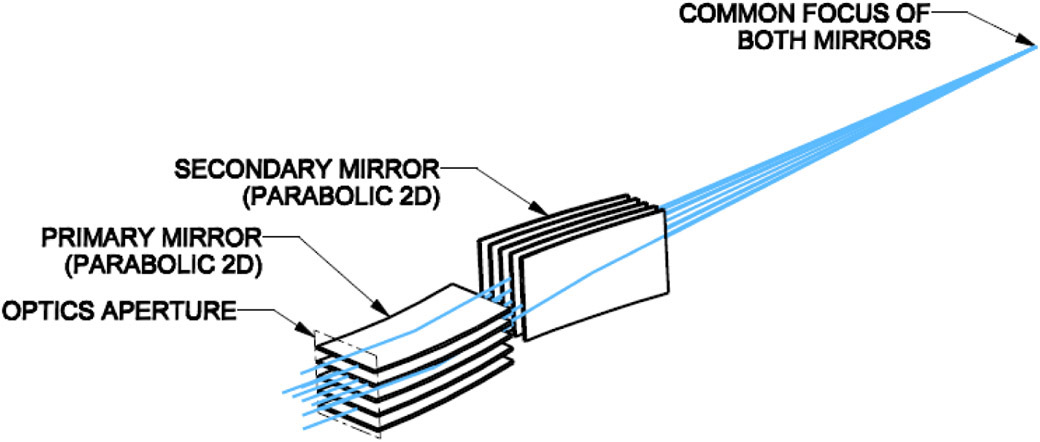}
   \end{tabular}
   \end{center}
   \caption[KB] 
%>>>> use \label inside caption to get Fig. number with \ref{}
   { \label{fig:KB} Kirkpatrick-Baez nested optical configuration (from \cite{KBfigref})}
   \end{figure}

 An alternative optical configuration is the Kirkpatrick-Baez (KB) geometry.  This geometry, originally proposed for X-rays before the Wolter configuration\cite{KB1948}, comprises two reflections where focusing in one plane is decoupled from focusing in a perpendicular plane. Here, both reflections are from parabolic mirror segments and nesting is straightforward as shown in Figure \ref{fig:KB}.
 
  The principal attraction of the KB approach is that the individual reflectors are nearly flat and relatively easy to fabricate.The principal drawback, however, is that there is only one reflection in each plane and this leads to a focal length twice that of the Wolter-1 configuration. This doubling of focal length places practical limits on the applicability of KB optics in a space mission. 
  
\subsection{Requirements on Figure and Surface}
Irregularities in the surface of a mirror will cause rays to be deflected thereby degrading the X-ray performance. Low frequency irregularities contribute to the overall geometrical figure accuracy of the mirror, whereas high frequency irregularities affect the surface roughness which can produce energy-dependent scattering. Typical figure accuracies at mid spatial scales (10 mm) has to be at the 0.1 micron level for few-arcsec-level optics. For efficient specular reflection of X rays, the surface roughness has to be smooth on the scale of the wavelength of the radiation. Thus, the requirement for the surfaces of X-ray optics is orders of magnitude more stringent than for optics operating in the visible region. The amount of flux scattered can be calculated from:
\begin{equation}
    \frac{I_s}{I_o} = 1-{\rm e}^{-\left(\frac{4 \pi \sigma \sin{\alpha}}{\lambda}\right)^2}
\end{equation}
Where $I_s$ is the scattered X-ray intensity, $I_o$ is the incident X-ray intensity, $\sigma$ is the rms surface roughness, $\alpha$ is the graze angle and $\lambda$ is the X-ray wavelength.  For a 6 keV photon ($\lambda=0.2$ nm), at an incidence angle of 0.5$^{\circ}$, a surface roughness of just 0.5 nm rms will cause scattering of approximately 8\% of incident flux. 

The above considerations place stringent requirements on figure and surface that can be challenging to meet, particularly given the very small graze angles involved with X-ray optics. The effective area of an X-ray mirror is the projected area multiplied by the reflectivity squared (for a typical Wolter-1 2-segment system). For a graze angle of $\alpha$, the projected area $=  2A / \sin\alpha$, where $A$ is the total surface area and the factor of two arises as there are two segments. Thus, for a graze angle of 0.5 degrees, the surface area that must be figured and polished will be $>$ 200 times the projected area for each mirror. This can be particularly challenging, inside highly-curved mirror shells. In addition, as will be shown later, the need for high throughput means that mirror shells are typically nested to increase collecting area and so this can place a premium on lighter-weight and thinner substrates for mirror shells that can be efficiently inserted and mounted. 

\subsection{Trades in Mirror Fabrication Approaches}
\begin{figure} [ht]
   \begin{center}
   \begin{tabular}{l} %% tabular useful for creating an array of images 
   \graphicspath{ {./images/} }
   \includegraphics[width=11.0cm]{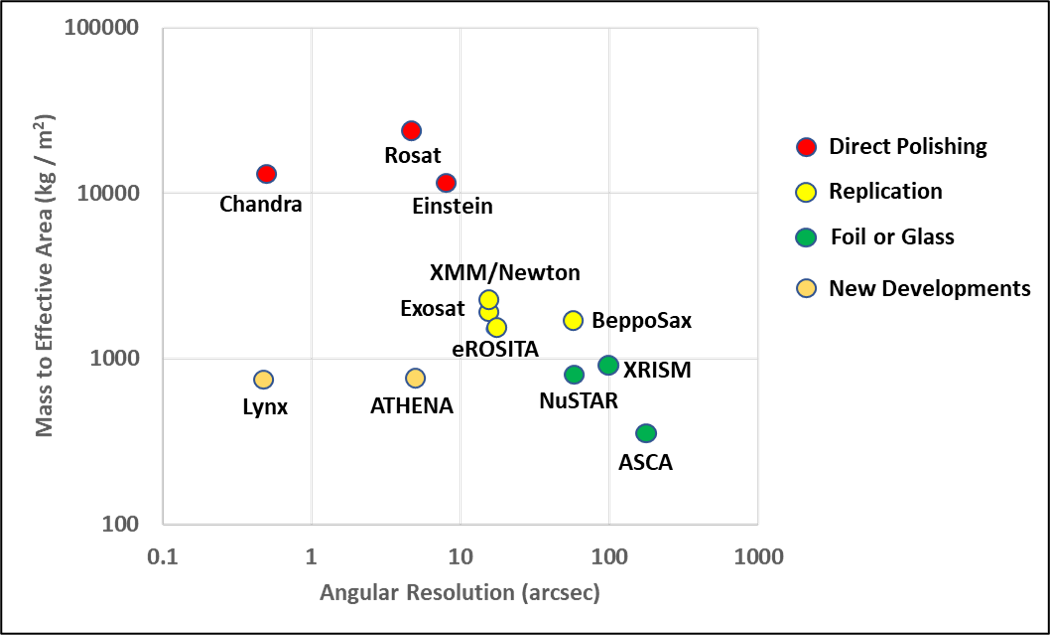}
   \end{tabular}
   \end{center}
   \caption[ResMass] 
%>>>> use \label inside caption to get Fig. number with \ref{}
   { \label{fig:ResMass} Areal mass vs angular resolution for various missions}
   \end{figure}

Developing a mirror for a specific application involves an optimization that trades angular resolution on the one hand for mass, ease of fabrication and cost on the other (see Figure \ref{fig:ResMass}). That is, the highest angular resolution X-ray mirror systems made to date for observatories that concentrate on imaging involve labor-intensive meticulous figuring and polishing of very thick, stable substrates that can be mounted without distortions. At the other extreme, thin foil optics, using coated aluminum reflectors, are relatively easy to fabricate, have large throughput (necessary for spectroscopy) and very low mass, but have relatively poor few-arcminute-level angular resolution.  Some examples of these various fabrication techniques are given in the next section. Figure \ref{fig:ResMass} shows the spectrum of HPD vs mass for X-ray optics fabricated and flown on a selection of missions to date, showing the traditional trade-off between mass and resolution. New missions seek to counter this dependency. Note that angular resolution is typically referred to in terms of Half-Power Diameter (HPD) or Half-Energy Width (HEW), rather than a more traditional Full-Width at Half Maximum (FWHM) measure. FWHM can be misleading as an apparently-high-resolution optic may have only have a very small fraction of the reflected flux within this diameter, the rest appearing in broad scattering wings of the point spread function, whereas the HPD/HEW has, by definition, half of the reflected rays within it.

%\end{document}

\section{Overview of Fabrication Techniques}

%\begin{document}
\subsection{Full Shell Optics}
\subsubsection{Direct}
Direct fabrication of full shell mirrors via grinding and polishing is a process that traces its ancestry to the earliest optical telescopes. The approach that has produced the highest resolution flight X-ray optics involves the fabrication of Zerodur ceramic shells, a few-cm thick to avoid fabrication and mounting-induced distortions, which are ground, figured, polished to 0.5 nm or less and coated inside with a highly reflective material such as gold or iridium. Optics of this type have been flown on multiple missions (e.g. Einstein, \S\ref{1970s}) and represent the traditional approach to mirror fabrication involving meticulous figuring and polishing of thick substrates.  The resulting mirrors can deliver sub-arcsecond angular resolution (see Chandra, \S\ref{1990s}) but have high cost (labor-intensive) and significant weight for the mirrors and their support structure. Other technologies have been developed that trade angular resolution for light weight, higher throughput (as the mirror shells can be nested more closely) and lower cost.

\subsubsection{Replication}
Replication seeks to transfer the figure of a precision master on to the surface of a complimentary mirror and has several distinct advantages. First, the master, usually termed a mandrel, can be relatively ‘massive’ and stable and so can be polished and figured without distortions. Second, it is somewhat easier to figure and polish the outside of cylinder, rather than the inside. Third, the mandrel can be used to fabricate multiple copies of the mirror leading to significant cost advantages if multiple mirror modules are being fabricated. The principal disadvantage is that the replicas do not typically exactly conform to the shape of the master, and so the resulting mirrors have poorer angular resolution (15-25 arcsec HPD) than those produced by direct fabrication. 
Although full-shell replication has been done using carriers slid over mandrels, with epoxy to take up the difference (e.g. EXOSAT, \S\ref{1980s}), by far the most full-shell replicated optics have been made using electroformed nickel. In this process, thin mirror shells are electroformed onto precisely figured mandrels from which they are later released through differential thermal contraction. Early optics for X-ray astronomy were fabricated using this process\cite{hudec2017replicated} and it continues to be in use today (e.g. XMM/Newton, \S\ref{1990s}) with stronger nickel cobalt alloys substituted for pure nickel (e.g. IXPE, \S\ref{2020s}).

\subsection{Segmented Optics}
Segmented optics split a single mirror shell into multiple components for both the parabolic and hyperbolic segments. The attraction of this process is that by splitting a system into many smaller segments much lighter reflectors can be used and these can be nested densely to improve throughput. To date multiple missions have flown aluminum foil reflectors (e.g. ASCA, \S\ref{1990s}), typically coated via epoxy replication off a smooth mandrel to improve surface roughness, which could be mass produced at relatively low cost. These are held in place by precision alignment bars that permit individual reflector to be inserted with no additional alignment necessary. In this way, extremely-low-cost, very-light-weight mirror modules can be developed, that trade angular resolution (typical foil optics are a few arcminute angular resolution) for cost and throughput.

An alternative approach to aluminum reflectors is to use slumped glass. The attraction of the glass is that it is available with very smooth surfaces due to its fabrication process. Certain glass types, such as low-expansion-coefficient float glasses, is available with sub 0.5 nm surface roughness, making it ideal for X-ray reflectors. By heating the glass to high temperatures ( $\sim60$0 $^{\circ}$C) on a suitably-figured Pyrex mandrels, the substrate can be given the desired shape. The NuSTAR (see \S\ref{2010s}) mission makes use of this technology. 

%\end{document}
\section{Chronological List of Mission with X-Ray Optics}
%\begin{document}
The following section gives a chronological listing of missions that flew, or plan to fly,  grazing-incidence X-ray optics. Each listing describes the type of optics used and their performance parameters, but also includes, as appropriate, challenges overcome and significant astrophysical results. Particular attention is paid to groundbreaking missions which significantly advanced the state of art in mirror technology and/or opened new fields of sensitive study that were not possible before.Note that aside from very early pioneering work with solar imaging the missions described are for non-solar X-ray astronomy. 
\subsection{Early Days} \label{earlyDays}
\textbf{Rocket-borne experiments:} Probably the first successful use of grazing-incidence X-ray optics in astronomy was in 1965\footnote{An earlier flight by the same group in 1963 yielded poor-quality images. Significant improvements were made to the telescopes for this second flight.} when solar X-ray images were obtained using three small identical grazing incidence telescopes, each with an X-ray-film-equipped camera at their focus, which were carried aloft by an aerobe rocket\cite{giacconi1965solar}. The X-ray mirrors were fabricated from electroformed nickel, were just 76 mm in diameter and 150 mm long, and had a focal length of 820 mm. The effective area was $\sim$5 cm$^2$ for the complete payload, sufficient for producing the first focused X-ray images of the sun at around 1.5 keV and with a FWHM angular resolution of about 1 arcminute. Over the next few years several groups flew grazing-incidence optics on solar rocket experiments with steadily-improving angular resolution and collecting areas\cite{underwood1967glancing,vaiana1968xray}. 

Despite these early successes, the adoption of this full-shell mirror technology to extra-solar X-ray astronomy was still some way off, although in 1970, a small rocket payload was launched with a one-dimensional Kirkpatrick-Baez type mirror which looked for emission from the Virgo and Cygnus regions of the sky\cite{Gorenstein1971rocket}. The mirror consisted of a series of chromium-coated float-glass plates which have the distinct advantage of having high-quality surfaces that need no additional polishing. Curved slightly in one direction to form a set of parabolas with a common focus, the reflector assembly produced a line image with a field of view in the focusing direction of 2$^{\circ}$, while orthogonal to this it was collimator-limited to 9$^{\circ}$.The rocket payload scanned both sky regions along the focusing direction and placed upper limits on source emission in the soft-X-ray band\cite{Gorenstein1972virgo}. The payload was later upgraded to a two-dimensional Kirkpatrick-Baez array, again using chromium-coated float-glass plates. The final optic had an effective area of 264 cm$^2$ at 1/4 keV and an angular resolution of 3 arcmin\cite{Gorenstein1975rocket}. The payload flew first in March 1975 and detected X-ray emission from the Algol region in a 0.15 - 2 keV energy band\cite{Gorenstein1977algol}. 
\subsection{1970s} \label{1970s}
\textbf{Skylab:}  Following the successes of the early short rocket flights using X-ray optics for solar observations, the need for much longer observing times was obvious. An opportunity arose when NASA’s Marshall Space Flight Center proposed a small orbiting space station to be launched on a Saturn V rocket, whose main scientific instrument would be a solar-pointing platform design to house multiple instruments covering the ultraviolet and X-ray energy band \cite{giacconi2008secrets}. When the proposal was accepted a call was issued for development of the scientific instruments. The winning X-ray proposal featured a telescope consisting of two nested mirror shells with a novel transmission grating in front of the telescope to provide imaging spectroscope. The mirror design specification was superior to those previously flown by the same group, with larger collecting area, by virtue of its nested array of longer-focal-length mirrors and with greatly improved angular resolution that would be achieved by careful figuring and polishing. The final design featured a pair of confocal mirror shells of diameters 310 mm and 230 mm and focal length 2.1 m, giving a geometric collecting area of 42 cm$^2$. Each mirror shell was fabricated from beryllium coated with Kanigen, a nickel-phosphorous alloy that is very hard and can be polished to a high-quality surface finish.

Measurements of angular resolution at around 1.7 keV revealed a narrow core with a FWHM resolution of around 10 arcsec and broad wings from surface scattering. The half-power diameter of the telescope was measured to be under 2 arcminutes\cite{vaiana1974xray}. It is important to note that for high-contrast imaging of bright sources such as the sun, the total amount of flux in the core is slightly less important than for stellar astronomy where X-ray sources are typically very weak and detector backgrounds play an important role if the focal spot is large. Similarly, collecting area is less important, particularly at low-X-ray energies, where solar X-ray fluxes are considerable.
Skylab operated between May 1973 and February 1974. The highly-successful X-ray spectroscopic telescope provided the highest angular resolution and sensitivity of any solar instrument at that time and provided a detailed view of dynamic structure of the sun and significantly impacted theories of coronal heating.

\textbf{Einstein:}  The Einstein Observatory (HEAO-2) was really the first astronomical imaging X-ray telescope, with an optic and corresponding single-photon-counting focal plane detector, to be used in orbit for astronomy. A principal challenge for the Einstein mirror was to reduce the scattering that contributed to significant wings in the Skylab mirror response function. This was necessary as the relatively low fluxes from astronomical X-ray sources required more of the flux to be in the high-resolution core of the mirrors response. Reducing scatter mean achieving a higher surface finish, and it was decided that this could be achieved with glass substrates which could be polished more finely. Experiments with fused silica confirmed that the required surface roughness could be reached\cite{miller1978high}.

The contract for the Einstein mirrors was awarded to Perkins Elmer, utilizing a material called Zerodur, which is a near-zero-thermal expansion coefficient fused quartz material. The final configuration was a set of four nested parabolic/hyperbolic pairs ranging in diameter from 0.34 m to 0.58 m (see Table \ref{tab:einstein} for mirror module parameters). A key challenge was to provide metrology in a free state so that the mirror segments true shape could be measured without distortions imparted by mounting. This was overcome by floating the mirrors in a bath of mercury so that their figures could be measured without gravitational distortions. 

\begin{table}[]
    \centering
    \caption{Einstein flight mirror configuration}
    \label{tab:einstein}
    \begin{tabular}{l|l}
    Parameter & Value \\
    \hline
    \hline
    Parabolic/hyperbolic pairs   & 4 \\
    Focal length  & 3.44 m\\
    Intersection diameters    & 0.34 - 0.58 m \\
    Graze angles & 42 - 72 arcminutes \\
    Segment length  & 0.51 m \\
    Material & Fused quartz \\
    Coating & Nickel \\
    Effective area & 400 cm$^2$ at 0.25 keV \\
    Angular resolution (HEW) & 8 arcsec at 0.28 keV \\
    \end{tabular}
\end{table}

Additional challenges included mounting the optic using materials with adequate rigidity yet low thermal expansion coefficient to avoid distortions in the test, launch and on-orbit operating environments. A special Invar alloy, heat treated to match the Zerodur expansion coefficient, was used for the interface to the mirrors at their ends and a graphite structure, also with matched expansion coefficient, tied the ends together\cite{miller1978high}.

To align and assemble the individual mirrors, each was hung by approximately 50 individual wires to reduce gravity-induced distortions. These supports had precisely-calculated off-load values to maintain the natural mirror shape during bonding. The alignment system was mounted on a five-axis micro-positioning stage to enable precise placement with respect to its optic pair, measured via optical monitoring. When the required accuracy was achieved the mirror was bonded to its Invar support system via injection of a suitable epoxy\cite{miller1978high}.

\begin{figure} [ht]
   \begin{center}
   \begin{tabular}{l} %% tabular useful for creating an array of images 
   \graphicspath{ {./images/} }
   \includegraphics[width=7.0cm]{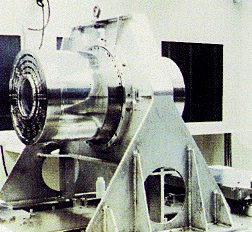}
   \end{tabular}
 %  \end{center}
   \caption[Einstein] 
%>>>> use \label inside caption to get Fig. number with \ref{}
   { \label{fig:einstein} The Einstein flight mirror assembly}
   \end{center}
   \end{figure}

\begin{figure} [ht]
   \begin{center}
   \begin{tabular}{l} %% tabular useful for creating an array of images 
   \graphicspath{ {./images/} }
   \includegraphics[width=6.0cm]{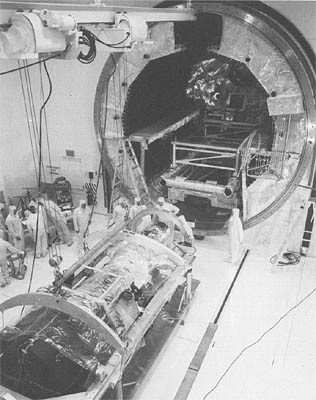}
   \end{tabular}
   \end{center}
   \caption[einsteinXRCF] 
%>>>> use \label inside caption to get Fig. number with \ref{}
   { \label{fig:einsteinXRCF} The Einstein Observatory at the MSFC X-ray calibration facility}
   \end{figure}

The Einstein mirror assembly (Figure \ref{fig:einstein}) was integrated with the flight instruments and shipped to Marshall Space Flight Center where a special calibration facility had been built. This facility had a 300-m long beam tube, to give low beam divergence, followed by a large instrument chamber of diameter 7 m and length 13 m (see Figure \ref{fig:einsteinXRCF}). 

X-ray measurements revealed that the telescope effective area agreed very well with pre-calibration predictions. Measurements of angular resolution showed a core in the point spread function that agreed well with predictions from axial figure and alignment errors. Moving out from the core, energy-dependent scattering, due to microroughness was observed at a slightly higher level than expected. Nevertheless, a half-energy width of around 8 arcsec was measured at 1.5 keV, with a FWHM core of under 4 arcsec. Measurements made in flight agreed very well with these calibration data\cite{vanspeybroeck1979einstein}.

Launched in November, 1978, the on-orbit performance of Einstein showed a dramatic improvement in sensitivity over previous missions. When compared with non-focusing instruments, such as the collimated proportional counters on UHURU, the improvement was about 3 orders of magnitude so that a source about one millionth the brightness of Sco-X-1 could now be detected. The net result was to open up the field of X-ray astronomy, broadening the possible research from not just studies of bright accreting binaries, but to studies of a wide range of source types from hot coronae in stars through supernova remnants to gas in clusters of galaxies and distant AGN.
\subsection{1980s} \label{1980s}
\textbf{Salyut 7:} In 1981 a soft X-ray telescope, RT-4M, was launched on the Salyut 7 space station. The telescope consisted of two nested mirrors shells with a Wolter-1 geometry. The inner shell had a diameter of 13 cm and was coated with Au while the outer shell had a diameter of 24 cm and was Ni coated. Produced in Czechoslovakia, the mirrors were manufactured using a replication process in which mm-thick nickel shells were replicated on glass mandrels, coated with an electrically-conducting material, and then reinforced with a 5-15-mm-thick layer of casted resin.\cite{Hudec1984ASR},\cite{Hudec1986SPIE}. The resulting mirror shells had an overall length, parabolic plus hyperbolic section, of 48 cm, and focal lengths of 42 cm, providing throughput in the range from 0.1 keV - 0.25 keV for the outer shell and 0.4 keV - 1 keV for the inner one. The resulting effective area was 90 cm$^2$ at 0.15 keV and 30 cm$^2$ at 0.5 keV. An angular resolution of 10 arcmin was obtained for the outer shell and 4 arcmin 
for the inner shell. 
RT-4M used a segmented proportional counter for a focal plane detector and this unfortunately failed within a week of launch and no science data were recorded.

\textbf{EXOSAT:}  The first European mission featuring X-ray optics was EXOSAT which was launched in May, 1983\cite{laine1979xray}. The optics for EXOSAT were fabricated using a novel type of replication technology. Specifically, a high-quality mandrel having the desired figure and surface roughness was fabricated from BK7 Schott glass and coated with  $\sim 90$ nm of evaporated gold. In parallel with this a high-quality mirror shell substrate was formed, identical in shape to the mandrel but approximately 30 microns larger in radius. During the replication process the substrate is brought over the mandrel and epoxy is injected between the two. When cured the shell is released through differential thermal expansion, taking the gold with it, and giving a mirror shell that replicates the high-quality figure and finish of the mandrel. For EXOSAT the mirror substrates were beryllium, 3.5 mm thick, and the resulting 2-shell mirror assemblies weighed just 7 kg, despite having an outer mirror diameter of 280 mm. The final payload consisted of two such modules with a focal length of 1.09 m giving an effective area of around 40 cm$^2$ each and a half-energy width of 10-15 arcsec, all at 0.8 keV\cite{dekorte1981exosat}.
\subsection{1990s} \label{1990s}
\textbf{ROSAT:}  The ROSAT mission was a joint German, US and British mission, proposed within the German advisory board as far back as 1975\cite{trumper1982rosat}. It was aimed at being a follow up to the Einstein mission and its main objective was high resolution imaging in the soft X-ray band from 0.1 to 2 keV where it conducted the first comprehensive all sky survey. It was launched in June of 1990 and was only planned to be operational for 18 months but it served the Astrophysics community for over 8 years and stopped operating in February of 1999, less than a year before the CHANDRA mission was launched. ROSAT paved the way for more sensitive all sky surveys such as eROSITA on the SRG mission which has extended the energy range to 10 keV and with much higher sensitivity.  

\begin{figure} [ht]
   \begin{center}
   \begin{tabular}{l} %% tabular useful for creating an array of images 
   \graphicspath{ {./images/} }
   \includegraphics[width=7.0cm]{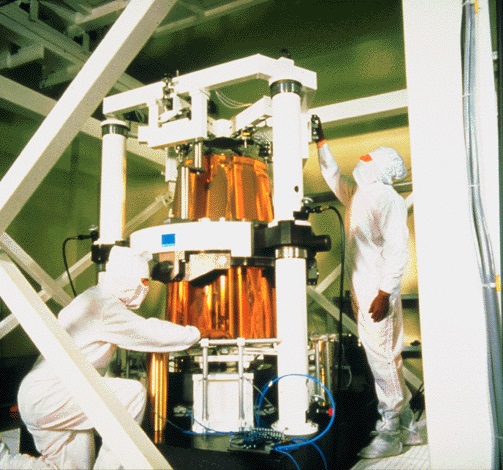}
   \end{tabular}
   \end{center}
   \caption[rosat] 
%>>>> use \label inside caption to get Fig. number with \ref{}
   { \label{fig:rosat} Rosat mirror integration}
   \end{figure}

The main instrument on ROSAT was a high resolution X-ray telescope with 2.4 m focal length, an inner mirror diameter of 37 cm and an outer diameter of 84 cm. It contained 4 nested shells and was built by Carl Zeiss in Germany\cite{aschenbach1988design}. The Half Power Diameter was 5 arcsec which has only been surpassed by CHANDRA. It carried two detectors:  a German position-sensitive proportional counter with a spatial resolution of 125 micron and a U.S.-provided high-resolution imager with a position resolution of 25 micron. In addition it carried a coaligned extreme XUV telescope with an associated wide field camera delivered by the UK. The energy range of this was from 0.042 keV to 0.21 keV. The X-ray telescope was a proper Wolter I geometry and the full shell mirrors were made out of highly polished Zerodur with an unprecedented micro-roughness of only 0.25 nm. A reflecting layer of Au was subsequently vacuum deposited. The net result was 420 cm$^2$ effective area at 1 keV and 470 cm$^2$  at 0.28 keV (see Figure \ref{fig:rosat}). The mass of the telescope and the focal plane detectors was 1.3 ton.
State of the art pointing provided 6 arcsec post facto attitude determination. The mission\cite{aschenbach1991first} resulted in a catalog of more than 150000 sources from the six month all sky survey.

\textbf{BBXRT:}  At small graze angles, the hyperbola and parabola of the Wolter-1 optical prescription can be replaced by straight conic sections with acceptable imaging performance. These sections can be fabricated from thin bent aluminum-foil reflectors, which can be densely packed to give a very lightweight X-ray optic with high throughput. The fabrication of such foil optics is greatly simplified by the fact that each mirror segment can be made from the same annulus, with only the length changing for different radii\cite{petre1985conical}. This means that the reflectors can be mass produced at very low cost. A precision-machined series of alignment/support bars hold the reflectors in place, with just enough free play to allow easy insertion of the foils. Very high packing fractions are possible for optimum on-axis performance.

The first foil mirror to fly was the SXS payload designed to look for X-ray emission from Supernova 1987A. This mirror had 68 concentrically nested mirror shells formed from 544 segments and was launched on a rocket in 1988. Issues with the pointing system prevented it from seeing the supernova, but it was followed by the Broad Band X-Ray Telescope which flew on the shuttle in December 1990\cite{petre2010xray}.
The BBXRT payload featured 2 identical mirror modules of 3.8 m focal length each with a pixelated silicon detector at its focus. The mirror modules each had 118 nested shells formed from nearly 1000 reflectors of 125 µm-thick aluminum. Each reflector was dipped in acrylic lacquer to provide a 10-µm-thick smooth layer which was then overcoated with 50 nm of evaporated gold to provide response up to 12 keV\cite{serlemitsos1985broad}. The use of the ultra-thin foils resulted in extremely light-weight optical assemblies, with the BBXRT mirror modules weighing just 20 kg each, despite having an effective area of nearly 300 cm$^2$ at 1 keV. On-orbit angular resolution was measured to be around 3 arcmin half-energy width.

BBXRT was flown as part of the Astro-1 payload aboard the space shuttle and successfully observed 85 different targets over the 9-day mission.

\textbf{ASCA:}  The ASCA mission was the first satellite mission to make use of the thin-foil X-ray technology, already flown BBXRT. It was also the first mission to make use of CCD detectors enabling high resolution broad band spectroscopic imaging of celestial X-ray sources for the first time. In this regard it was the perfect complementary match to the ROSAT mission at the time. ASCA was launched in February 1993 and stopped operating July 2000. It carried 4 X-ray telescopes – two of which had gas detectors at the focal plane and two had CCD cameras. The mission was a collaboration between Japan and the US.

\begin{figure} [ht]
   \begin{center}
   \begin{tabular}{l} %% tabular useful for creating an array of images 
   \graphicspath{ {./images/} }
   \includegraphics[width=7.0cm]{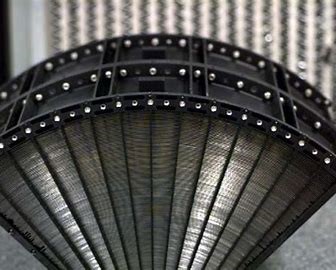}
   \end{tabular}
   \end{center}
   \caption[asca] 
%>>>> use \label inside caption to get Fig. number with \ref{}
   { \label{fig:asca} A segment of the ASCA optic with 120 nested foil shells}
   \end{figure}

The four ASCA X-ray telescopes were supplied by Goddard Spaceflight Center and as with BBXRT were conical approximations to a Wolter I geometry. The reflectors were made of thin Al foils – 127 micron in thickness and were dip-lacquered to smooth short-length-scale roughness up to $\sim100$ micron\cite{serlemitsos1985conical}. The thickness of the lacquer was 10-20 microns and a 50 nm film of Au was evaporated on top as the reflecting layer. The resulting mirror shells were tightly nested from an inner diameter of 100 mm to an outer diameter of 345 mm. Each telescope had 120 nested shells and a focal length of 3.5 m (see Figure \ref{fig:asca}).

ASCA was launched in stowed configuration and extended after launch to achieve the 3.5 m focal length. The energy range extended up to 12 keV, the effective area was 1200 cm$^2$ at 1 keV and 600 cm$^2$ at 7 keV and the Field of View was 24 arcmin at 1 keV and 16 arcmin at 7 keV. An angular resolution of 2.9 arcmin was achieved, mainly limited by mm scale residual figure error of the rolled Al foils and the accuracy of mounting individual reflectors using the alignment bars\cite{serlemitsos1995xray}. The total mass of all 4 telescopes was just under 40 kg. Table \ref{tab:asca} details the ASCA mirror parameters.

\begin{table}[]
    \centering
    \caption{ASCA mirror parameters}
    \label{tab:asca}
    \begin{tabular}{l|l}
Parameter & Value \\
    \hline
    \hline
    Number of telescopes & 4 \\
    Number of nested segments   & 120 \\
    Focal length  & 3.44 m\\
    Intersection diameters   & 0.34 - 0.58 m \\
    Graze angles & 42 - 72 arcminutes \\
    Segment length  & 0.51 m \\
    Material & Aluminum \\
    Coating & Acrylic-adhered gold \\
    Effective area & 1300 / 600 cm$^2$ at 1 / 7 keV \\
    Angular resolution (HEW) & 2.9 arcmin \\
    \end{tabular}
\end{table}

\textbf{Beppo-SAX:}  was an Italian-Dutch mission launched in 1996 and deactivated 2002. It carried 4 focusing telescopes with a focal length of 1.85 m and a useful energy response from 0.1-10 keV.  In addition, it carried coded mask instruments with response up to 300 keV. The optics of choice was thin (0.2 – 0.4 mm) electroformed Ni shells in a conical approximation to a Wolter I optic with thirty shells in each telescope\cite{citterio1986optics}. A moderate resolution of 1 arcmin was required dictated by the resolution of the focal plane detectors. The combined effective area at 7 keV was 175 cm$^2$. Beppo-SAX played a significant role in resolving the origin of gamma-ray bursts\cite{piro2013beppo}.

\textbf{Chandra (AXAF):}  The foundations for Chandra were laid in a proposal\cite{giacconi1963experimental}, ‘An experimental Program in X-Ray Astronomy’, led by Riccardo Giacconi and submitted to NASA in 1963 shortly after the original discovery of Sco-X-1 and the X-ray background. In this proposal a seven-phase program was described, starting with additional rocket flights and culminating with a 10-m-focal length observatory with a 1.2-m-diameter X-ray optical system\cite{giacconi2008secrets}. In 1976, an unsolicited proposal was submitted to NASA, again with Riccardo Giacconi as the Principal Investigator, to study this large X-ray telescope which would have  sub-arcsecond angular resolution, very high sensitivity by virtues of an effective area greater than 1000 cm$^2$, and a planned lifetime of order 10 years. Even though the Einstein observatory was yet to be launched (see section \ref{1970s}) , NASA approved the study and gave the Management role of what was then named the Advanced X-ray Astrophysics Facility (AXAF) to the Marshall Space Flight Center with the Smithsonian Astrophysical Observatory (SAO) providing scientific and technical support\cite{tananbaum2019chandra}.

The 1980’s saw a period of technical studies, the most important of these involved the development of the X-ray mirrors. As part of this a Technical Mirror Assembly (TMA) was built to demonstrate that the required angular resolution and overall imaging performance, an order of magnitude better than achieved with Einstein, could be met. The TMA was a 2/3-size replica of the innermost pair of the then baseline design for AXAF, with a diameter of 0.4 m and a focal length of 6 m, which could be X-ray tested in the existing facility at MSFC. Several commercial manufacturers were funded to provide the TMA, but in the end only Perkin-Elmer succeeded, delivering an optic that satisfied AXAF requirements with 90\% encircled energy measure within a radius of 1 arc second.
Although the TMA was a success and AXAF had been rated as the top priority by the 1980 astrophysics decadal survey, there were still many hurdles to overcome. NASA submitted a request for a formal start to the program for 1989, but was turned down by the office of management and budget. Finally, due in part to lingering concerns with Hubble Space Telescope Optics, funds were given to demonstrate that, within 3 years, the largest mirror pair (of 6 nested pairs in the then baseline design) could be built and would meet specifications as demonstrated through X-ray testing. This fabrication work would be carried out by Hughes Danbury who now owned the optics division of Perkin Elmer that had manufactured the TMA. In parallel with the mirror work, the X-ray calibration facility at MSFC was expanded to accommodate the 10 m AXAF focal length, both through extension of the vacuum beamline to over 500 m and through an expansion of the instrument chamber.

\begin{figure} [ht]
   \begin{center}
   \begin{tabular}{l} %% tabular useful for creating an array of images 
   \graphicspath{ {./images/} }
   \includegraphics[width=7.0cm]{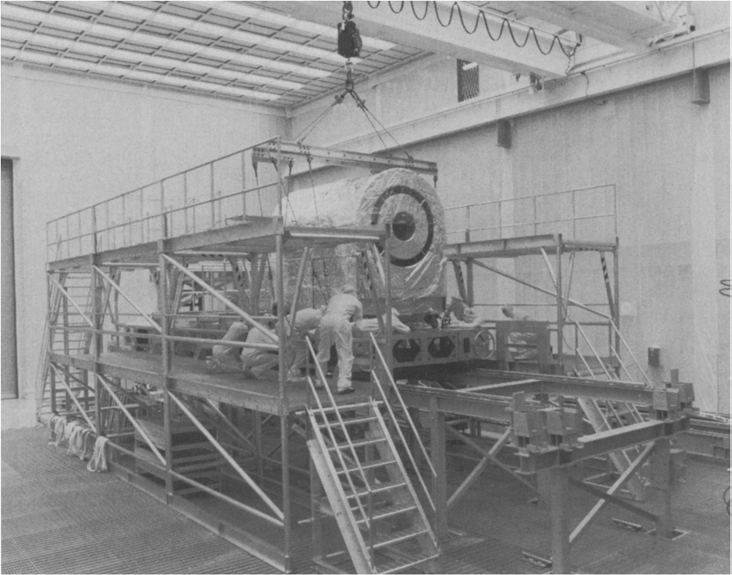}
   \end{tabular}
   \end{center}
   \caption[axafVETA] 
%>>>> use \label inside caption to get Fig. number with \ref{}
   { \label{fig:axafVETA} The AXAF Verification Engineering Test Article (VETA) being tested at MSFC}
   \end{figure}

Thus, a Verification Engineering Test Article was fabricated to develop and test the technology for optics fabrication and, equally important, to develop the necessary metrology. VETA was the uncoated outermost parabolic/hyperbolic shell and was measured at the newly upgraded XRCF (see Figure \ref{fig:axafVETA}). After correcting for gravitational distortions, the mirror was found to have an angular resolution of ¼ arcsec (Full Width at Half Maximum) and a half-power diameter of under 1 arcsec, meeting project requirements\cite{weisskopf1993axaf}.

A final major hurdle remained before flight hardware would be built. In 1992 it became evident that the new NASA budget could not support the projected costs of AXAF and that it would need to be descoped or cancelled. After more months of study AXAF was split into 2 missions - an imaging mission with a reduced number of high-resolution X-ray optics, two imaging instruments, plus gratings (AXAF-I) and a spectroscopic mission with poorer angular resolution ($\sim 1$ arcminute) and an X-ray calorimeter (AXAF-S). About a year later the AXAF-S mission was cancelled.

A final round of studies came up with the current (Chandra) configuration for AXAF, with four nested pairs of mirrors. A decision was made to place AXAF in a highly-elliptical, high-earth orbit, which somewhat compensated for the loss of collecting area (going from 6 mirror pairs to 4) by giving much higher viewing efficiencies. The high-earth orbit also meant that a longer lifetime could be expected without the continual day/night thermal-stress-inducing transition of low-earth orbit. Finally, in 1992, the project could complete the design and start fabricating the AXAF Observatory.

The final flight mirror configuration consists of 4 nested parabolic hyperbolic pairs with parameters given in Table \ref{tab:chandra}.

\begin{table}[]
    \centering
    \caption{AXAF (Chandra) flight mirror configuration}
    \label{tab:chandra}
    \begin{tabular}{l|l}
Parameter & Value \\
    \hline
    \hline
    Number of nested segments   & 4 \\
    Focal length  & 10 m\\
    Intersection diameters    & 0.63, 0.85, 0.97, 1.2 m \\
    Graze angles & 27 37 42 52 arcminutes \\
    Segment length  & 0.85 m \\
    Wall Thickness & 16 - 24 mm \\
    Material & Zerodur \\
    Coating & Iriduim, 10 nm \\
    Effective area & 770 cm$^2$ at 1 keV \\
    Angular resolution (HEW) & 0.5 arcsec \\
    \end{tabular}
\end{table}

\begin{figure} [ht]
   \begin{center}
   \begin{tabular}{l} %% tabular useful for creating an array of images 
   \graphicspath{ {./images/} }
   \includegraphics[width=7.0cm]{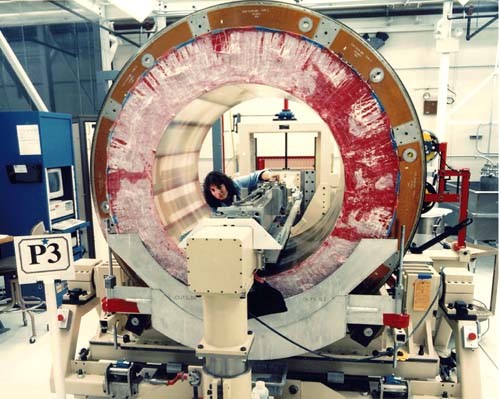}
   \end{tabular}
   \end{center}
   \caption[chandraGrinding] 
%>>>> use \label inside caption to get Fig. number with \ref{}
   { \label{fig:chandraGrinding} One of the AXAF paraboloid mirrors during grinding}
   \end{figure}

The four flight mirror pairs went through identical fabrication phases which consisted of coarse and fine grinding (see Figure \ref{fig:chandraGrinding}), followed by polishing and then a final smoothing run. Typically, these were carried out with computer-controlled small tools and an iterative process where metrology was taken, compared with requirement and a hit map produced showing where and how much material to remove in the next run. The process would converge and then move to the next fabrication phase. Metrology equipment, both contact and non-contact was used to determine axial figure errors, circularity and surface roughness. Figure measurements were performed in temperature-controlled environments with the optic oriented vertically to prevent gravitational sag. Typically, there were four grinding and four polishing cycles carried out for each element. The final surface smoothing was carried out with a large lap to avoid disturbing the mirror axial figure profile. A surface roughness over the central 90\% of all optical elements was in the range 0.19 – 0.35 nm over the micron to mm spatial scale\cite{weisskopf2012chandra}, a very low number that ensures that a large amount of flux remains within the core of the mirror’s point spread function.

\begin{figure} [ht]
   \begin{center}
   \begin{tabular}{l} %% tabular useful for creating an array of images 
   \graphicspath{ {./images/} }
   \includegraphics[width=8.0cm]{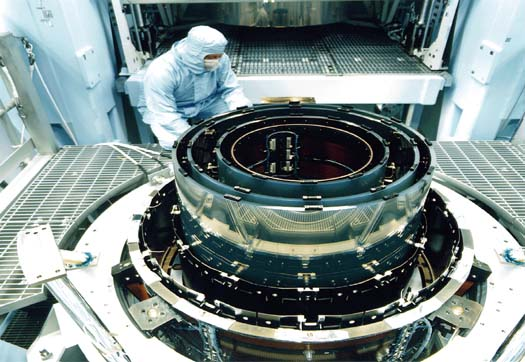}
   \end{tabular}
   \end{center}
   \caption[chandraAlignment] 
%>>>> use \label inside caption to get Fig. number with \ref{}
   { \label{fig:chandraAlignment} AXAF mirror module assembly}
   \end{figure}

Alignment and assembly of the mirror elements was carried out by Eastman Kodak Company. Each mirror was supported by flexures which were bonded near the middle of the optic. To bond the flexures the mirrors were oriented with the optical axis vertical in a near-strain-free state, positioned mechanically and monitored optically using a laser system which passed through the optic and back on itself using an auto-collimating flat (Figure \ref{fig:chandraAlignment} shows the AXAF mirror assembly).

The mirror assembly was taken to MSFC's X-ray calibration facility in December 1996 for a full end-to-end calibration with the flight instrumentation over a 6-month period working round the clock – the most comprehensive calibration ever for an X-ray mission. These calibrations confirmed the sub-arc-sec imaging performance of the X-ray optic assembly.

\begin{figure} [ht]
   \begin{center}
   \begin{tabular}{l} %% tabular useful for creating an array of images 
   \graphicspath{ {./images/} }
   \includegraphics[width=6.0cm]{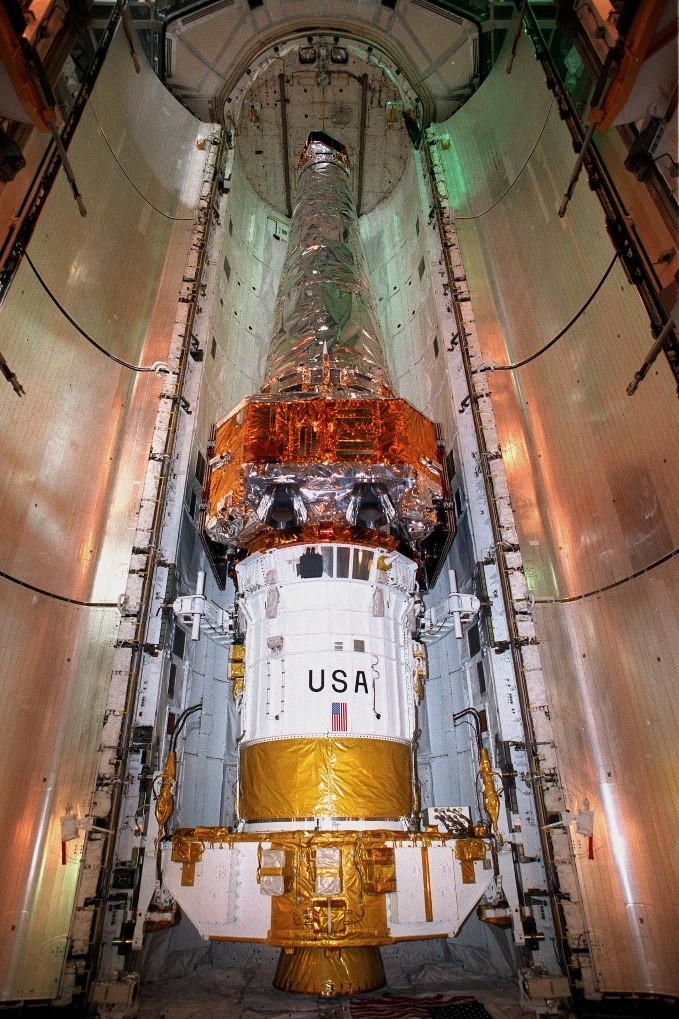}
   \end{tabular}
   \end{center}
   \caption[chandraLaunch] 
%>>>> use \label inside caption to get Fig. number with \ref{}
   { \label{fig:chandraLaunch} The Chandra Observatory in the shuttle cargo bay}
   \end{figure}

AXAF, renamed Chandra after a worldwide competition, was launched aboard the space shuttle on July 23, 1999 (see Figure \ref{fig:chandraLaunch}). Once in low-earth orbit, and released from the shuttle, an inertial upper stage was fired to place the observatory into a highly elliptical earth orbit and then via firings from the spacecraft engines into its final orbit of 9,700 km x 139, 000 km. First light for the observatory was Aug 12, 1999 and then on Aug 19, an image of the supernova remnant CAS A revealed a wealth of detailed information included the original star, the original source of the explosion. Among its many accomplishments, Chandra has since observed X-ray emission from all classes of normal stars, provided evidence for feedback mechanism between supermassive black holes and their surrounds that modulates their X-ray emission and, working in concert with optical observatories, has provided compelling evidence for the existence of dark matter, which makes up nearly 30\% of the mass of the Universe. After 22 years of operation, Chandra continues to provide groundbreaking science coupling very high sensitivity (10 orders of magnitude dynamic range) with unprecedented angular resolution.

\textbf{XMM/NEWTON:}  Europe’s first dedicated X-ray astronomy mission EXOSAT (see section \ref{1980s}) was the ESA equivalent of NASA’s Einstein mission and it carried two low energy X-ray telescopes (\cite{lumb2012xmm}). It was operational for 3 years from 1983 to 1986 and made nearly 2000 observations. Already before its launch it was clear that a much more ambitious spectroscopy mission in the X-ray energy range held great scientific promise. An ambitious mission proposal was formulated at this stage\cite{jansen2001xmm} and underwent significant design changes over the next 5-year period. The mission was initially called the XMM (X-ray Multimirror Mission) and was aptly called so as the first mission design called for 12 low energy telescopes and 7 high energy telescopes.  The mission became the second corner stone mission of the Horizon 2000 program. It is to date the largest scientific mission built in Europe. The first set of comprehensive mission objectives was presented at a meeting in Denmark in 1985. Over the next 2 years the telescope working group reduced the number of telescopes to 7 units which eventually became 3 telescopes operating at a focal length of 7.5 meters and covering the energy range from 0.1 to 12 keV with an unprecedented effective area and wide field of view. The complete set of instruments was selected by 1989. With a focus on spectroscopy and large effective area the required imaging resolution was around 15 arcsec\cite{aschenbach2000imaging}. This would be the perfect complementary match to the CHANDRA mission being defined at same time, which ultimately would emphasize sub-arcsecond imaging. 

\begin{figure} [ht]
   \begin{center}
   \begin{tabular}{l} %% tabular useful for creating an array of images 
   \graphicspath{ {./images/} }
   \includegraphics[width=8.0cm]{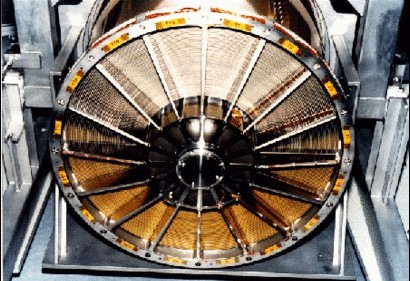}
   \end{tabular}
   \end{center}
   \caption[xmm] 
%>>>> use \label inside caption to get Fig. number with \ref{}
   { \label{fig:xmm} An XMM/Newton mirror module with 58 nested nickel shells}
   \end{figure}

Realizing XMM/Newton goals was however no trivial task. It was clear that a highly nested set of mirrors was required to reach the effective area requirements and the solution that was eventually implemented was full revolution replicated nickel shells. Each telescope consists of 58 cylindrical, nested Wolter-1-type mirrors (see Figure \ref{fig:xmm}) developed by Media Lario of Italy, each 600 mm long and ranging in diameter from 306 to 700 mm , producing a total collecting area of 4,425 cm$^2$  at 1.5 keV and 1,740 cm$^2$  at 8 keV\cite{gondoin1998calibration}. The mirrors range from 0.47 mm thick for the innermost mirror to 1.07 mm thick for the outermost. Each mirror was built by vapour-depositing a 250 nm layer of gold reflecting surface onto a highly polished aluminium mandrel, followed by electroforming a monolithic nickel shell onto the gold. The finished mirrors were glued into the grooves of an inconel spider, which keeps them aligned to within the five-micron tolerance required to achieve adequate angular resolution. The shells were only held at one end to avoid distortion from eventual over-constraining. The mandrels were manufactured by Carl Zeiss AG, and the electroforming and final assembly were performed by Media Lario with contributions from  Kayser Threde\cite{dechambure1997}.

The construction started in 1995 and the three telescopes were completed in December of 1998 and subsequently underwent calibration. The telescope modules are only supported at one to avoid distortions so special care needed to be taken during calibration. Calibration took place at the Panter long beam facility in Munich\cite{gondoin1998calibration} and additional calibration data was taken at a special vertical beam facility that was constructed for this purpose\cite{tock1997calibration}. The on ground and eventual on-orbit calibration was in very good agreement. The telescope performance parameters resulting from the calibration are given below. 

\begin{table}[]
    \centering
    \caption{XMM/Newton telescope parameters}
    \label{tab:xmm}
    \begin{tabular}{l|l}
Parameter & Value \\
    \hline
    \hline
    Telescopes   & 3 \\
    Number of shells & 58 \\
    Focal length  & 7.5 m\\
    Intersection diameters    & 0.34 - 0.70 m \\
    Graze angles & 42 - 72 arcminutes \\
    %Shell Thickness & 0.47 - 1.07 mm \\
    Segment length  & 0.51 m \\
    Material & Nickel \\
    Coating & Gold \\
    Effective area & 4400 cm$^2$ at 1.5 keV \\
    Angular resolution (HEW) & 15 arcsec at 0.28 keV \\
    \end{tabular}
\end{table}

\begin{figure} [ht]
   \begin{center}
   \begin{tabular}{l} %% tabular useful for creating an array of images 
   \graphicspath{ {./images/} }
   \includegraphics[width=9.0cm]{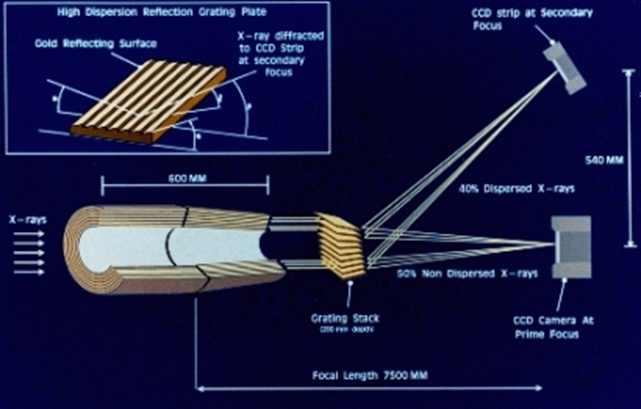}
   \end{tabular}
   \end{center}
   \caption[xmmGrating] 
%>>>> use \label inside caption to get Fig. number with \ref{}
   { \label{fig:xmmGrating} XMM/Newton diffraction grating configuration}
   \end{figure}

The focal plane detectors were wide-field-of-view CCD detectors covering the field of view of 30 arcmin and moderate energy resolution of $E/\Delta E$ of 20-50. In addition two of the telescopes are combined with a reflection grating spectrometer providing high-resolution spectroscopy.  The grating array is placed between the telescope and the focal plane and deflects 40\% of the focused beam (see Figure \ref{fig:xmmGrating}). The energy range of the reflection grating array is 0.35 keV – 2.5 keV covering the K shell lines of Oxygen, Neon, Magnesium, Aluminium and Silicon as well as the L lines of Iron.   The mission also carried an optical monitor providing simultaneous optical/UV observations of the X-ray field.

XMM was ready for launch in September of 1999 at which time the mission got its official name XMM-Newton. The launch was on an ARIANE V rocket which placed the mission in a highly elliptical orbit with an inclination angle close to 40 degrees.

XMM-Newton's superior effective area allows for spectroscopic studies of fainter sources compared to Chandra. This has led to significant contributions in all areas of high-energy astronomy from solar system objects to some of the most distant super-massive black holes.Some highlights include: spectra of tidal disruption events showing gas flows from disrupted stars near to the center of supermassive black holes;studies of the dynamic behavior of the inner accretion disk and corona showing the fundamental interaction between radiation and matter at the heart of quasars using reverberation mapping technique and; evidence that a large fraction of the missing baryons reside in filaments of the cosmic web observed at the outskirts of massive clusters of galaxies.

\subsection{2000s} \label{2000s}
\textbf{SWIFT:}  The Neil Gehrels Swift Observatory, a NASA mid-size explorer launched in November 2004, carried a suite of instruments to explore gamma-ray burst. Among these was the Swift X-Ray Telescope (XRT)\cite{burrows2000swift} which was designed to measure the fluxes, spectra and light curves from the burst, including their afterglow phase. The XRT has a single mirror module, of 3.5 m focal length, comprised of 12 concentrically-nested mirror shells that were fabricated using the electroformed-nickel replication process. The shells range in diameter from 191 mm - 300 mm and give a total effective area of 120 cm$^2$ at 1.5 keV and an angular resolution of 15 arcsec HPD. The XRT mirror module is one of two originally fabricated for the JET-X telescope that was to fly on the later-cancelled Spectrum-X mission\cite{citterio1996characteristics}.

\textbf{SUZAKU:} The SUZAKU (formerly ASTRO-E2) mission was the follow up to The ASCA mission and like ASCA is a joint Japanese-US mission\cite{mitsuda2007x}. It was operational from 2005 to 2015 and carried five X-ray telescopes, four of which were dedicated to imaging and one dedicated to spectroscopy. The focal length, facilitated by an extendible bench, was 4.75 m for the imaging telescopes and 4.5 m for the spectroscopy telescope. The telescopes were conical approximations to a Wolter I geometry and based on thin segmented Al foils as was the case for ASCA but instead of smoothing the Al surface with a lacquer coating a novel replication technique was used where Au (thickness $\sim 1000 \AA$ ) was sputtered on a smooth glass mandrel and, via epoxy, replicated to the mirror foils. The inherent large-scale figure error of the glass mandrel and the accuracy of mounting of individual mirror segments limited an energy independent angular resolution to 2 arcmin\cite{serlemitsos2007x}. This was, however, a significant improvement over the ASCA mission optics. The energy independence of the imaging resolution is testament to the much smoother short length scale (less than $\sim50$ micron) roughness of the reflecting surface as opposed to the ASCA mission where an increased half power diameter with energy was observed due the short length scale roughness. The useful effective area range extended from 0.2 keV - 12 keV and was 450 cm$^2$ at 1.5 keV and 250 cm$^2$ at 7 keV. The science instruments at the focal plane were CCD cameras with a field of view of 20 arcmin for the imaging telescopes. The mission was the first to launch a high-resolution micro calorimeter at the focus of the spectroscopy telescope. It also carried collimated detector sensitive to 600 keV. Unfortunately, the micro calorimeter failed due to loss of cryogen before routine operations was possible.

\subsection{2010s} \label{2010s}
\textbf{NuSTAR:}  One of the remaining frontiers of high energy Astrophysics in the post Chandra/XMM era was to extend the energy band of focusing optics into the hard X-ray region previously only observed with non- focusing coded mask instruments suffering from high background and poor sensitivity. Several balloon mission campaigns InFocus\cite{berendse2003production}, HERO\cite{ramsey2002hero} and HEFT\cite{harrison2000development} were eventually realized to demonstrate the capability of doing this.  In the case of HEFT, a novel approach to high throughput optics with moderate resolution was the development of inherently smooth thermally formed glass substrates and a novel method of mounting these in a tightly nested geometry using graphite spacers precisely machined and epoxied between adjacent shells\cite{hailey1997substrates}. The mirrors was coated with novel designs of graded d-spacing W/Si and Pt/C multilayer coatings\cite{christensen2011coatings}. These enabled the reflection of hard X-rays at larger graze angles than could be achieved with simple metal coatings thus allowing for hard X-ray telescope designs with a practical focal length of 10 m. The K absorption edge of Pt limited the upper response to 80 keV. The novel accurate mounting scheme resulted in an imaging resolution of just under 1 arcmin , the best achieved to date in a complete telescope consisting of highly segmented thin mirror shells. 

Based on the HEFT balloon mission demonstrations, the NuSTAR (Nuclear Spectroscopic Telescope Array) mission concept was proposed to NASA as a SMEX mission in 2003, with Caltech as the the PI Institution. The mission also relied on independent development of an extendible boom to facilitate the 10 m focal length\cite{craig2011fabrication}. The optical design allowed for effective areas at hard X-rays up to 80 keV similar to what was achieved for soft X-rays around 1 keV on the first focusing X-ray mission Einstein. This in turn translated to an improvement of the sensitivity of 2-3 orders of magnitude essentially opening this part of the electromagnetic spectrum to sensitive studies of narrow field of view sources. 

\begin{figure} [ht]
   \begin{center}
   \begin{tabular}{l} %% tabular useful for creating an array of images 
   \graphicspath{ {./images/} }
   \includegraphics[width=7.0cm]{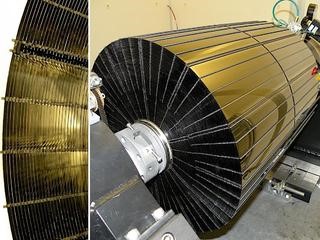}
   \end{tabular}
   \end{center}
   \caption[nustarMirror] 
%>>>> use \label inside caption to get Fig. number with \ref{}
   { \label{fig:nustarMirror} One of NuSTAR's flight mirror assemblies with 133 shells of thin (0.21 mm) glass}
   \end{figure}

\begin{figure} [ht]
   \begin{center}
   \begin{tabular}{l} %% tabular useful for creating an array of images 
   \graphicspath{ {./images/} }
   \includegraphics[width=8.0cm]{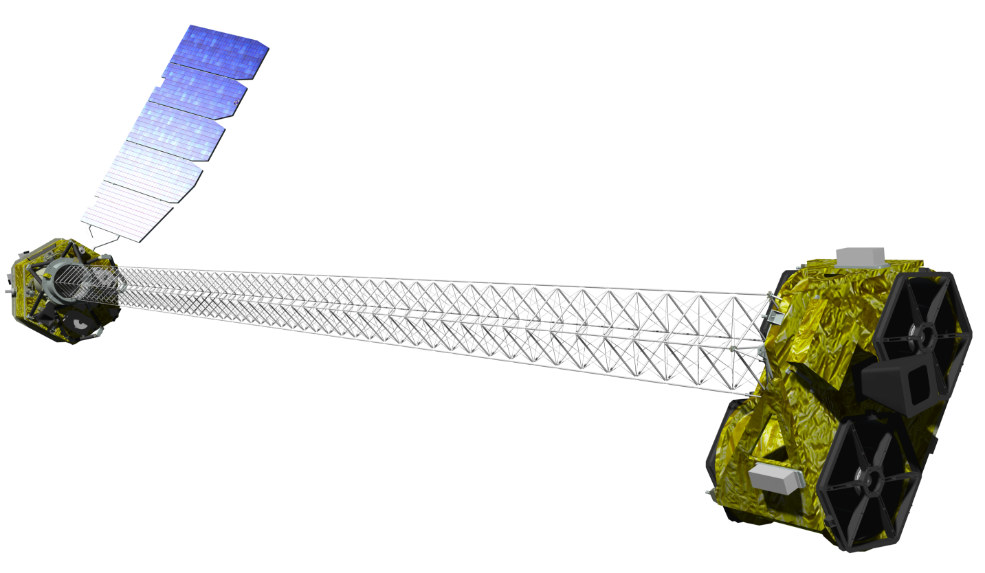}
   \end{tabular}
   \end{center}
   \caption[nustarObservatory] 
%>>>> use \label inside caption to get Fig. number with \ref{}
   { \label{fig:nustarObservatory} Artist's rendition of NuSTAR with two telescope modules, the extended boom and the detector platform}
   \end{figure}

NuSTAR was selected for flight in 2005. Due to budgetary constraints the mission was, however, cancelled in early 2006. In September of 2007 the mission was restarted and was launched on June 13 of 2012 on a Pegasus rocket into a near equatorial orbit as the eleventh SMEX mission. The mission baseline lifetime was two years but it is still in operation nine years after launch. NuSTAR carries two telescope modules (see Figures \ref{fig:nustarMirror} and \ref{fig:nustarObservatory}) that were built in the two year period from 2010-2011. In fact, 3 modules were produced. The first one was found to have significant issues with the reflecting coating (Pt/SiC). Another coating was developed (Pt/C) and the two subsequent modules met requirements and were selected for flight. The telescopes were calibrated at a custom facility at Colombia University\cite{brejnholt2012nustar}. The effective areas match very closely those of XMM-Newton at ~ 10 keV and thus lends itself to simultaneous observations with XMM- Newton and NuSTAR. The in-flight calibration of the two telescopes imaging and effective area performance confirmed the results of the on ground calibration apart from a small asymmetric broadening of the core of the PSF and effective area features around the L-edges of Pt and W\cite{madsen2015calibration}. NuSTAR telescope parameters are given in Table \ref{tab:nustar}.

\begin{table}[]
    \centering
    \caption{NuSTAR mirror module parameters}
    \label{tab:nustar}
    \begin{tabular}{l|l}
    Parameter & Value \\
    \hline
    \hline
    Telescopes   & 2 \\
    Focal length  & 10.15 m\\
    Intersection diameters    &  109 - 382 mm\\
    Wall thickness & 0.21 mm \\
    Graze angles & 4.5 - 16.2 arcmin \\
    Segment length  & 0.45 m \\
    Material & D263 Borosilicate glass \\
    Coating & W/Si and Pt/C graded d-spacing multilayers \\
    Mass per module & 37 kg \\
    Effective area &  800, 300 cm$^2$ at 10, 30 keV\\
    Angular resolution (HEW) & 58 arcsec \\
    \end{tabular}
\end{table}

The focal plane detectors for the NuSTAR mission is novel CdZnTe pixelated solid state detectors. The overall mission and the resulting science performance has been presented\cite{harrison2013nuclear}. The mission has observed a broad range of objects. Some of the most significant Science high lights are the observation of the broad-band spectrum of a supermassive black hole at the center of NGC 1365 permitting a determination of the spin state of the system\cite{risaliti2013rapidly} and the theoretical implications of how a supermassive black grows, as well as the observation of the asymmetric distribution of $^{44}$Ti emission near 80 keV in the CAS-A supernova remnant which is key to distinguishing between different models of how a supernova explosion actually occurs\cite{grefenstette2014asymmetries}.

\textbf{ASTROSAT:}  Astrosat, a multiwavelength observatory launched in October 2015 and still operational, was the first dedicated Indian Astronomy satellite. It carries a Soft X-ray Telescope based on thin (0.2 mm) segmented gold-coated Al foils in a nested 40 shell conical approximation to a Wolter I geometry\cite{singh2017soft}.
The focal length is 2 m and the focal plane detector is a CCD with a 40 armin Field of View. This restricts the effective area to the 0.3 - 8 keV range with a maximum effective area at 1 keV of 120 cm$^2$. The imaging resolution is modest at $\sim 5$ arcmin Half Power Radius.

\textbf{NICER:}  The Neutron-star Interior Composition ExploreR (NICER), launched in 2017, is an instrument on the International Space Station that combines 56 X-Ray Concentrators (XRC) with a corresponding array of silicon drift detectors. Each X-ray concentrator has 24 concentrically-nested shells, raging in diameters from 30 mm to 105 mm, with a focal length of 1.085 m\cite{okajima2016performance}. The shells are fabricated from aluminum foil that has been thermally shaped on individual aluminum forming mandrels and are full shells of revolution (minus a small gap) rather than multiple segments. A high-quality surface is subsequently achieved through an epoxy replication of the formed foils off glass mandrels. As the emphasis is on signal to noise rather than off-axis imaging, the XRCs consist of just a parabolic section rather than the combination of parabola and hyperbola. This significantly reduces mass and cost; the mass of each XRC is just 325 g. The use of heat forming means that a true parabolic shape can be imparted to the foils and this provides a significant improvement to the imaging performance (and also signal to noise) over the usual conic sections used in traditional foil optics\cite{balsamo2016shrink}.

\textbf{Spectrum-Röntgen-Gamma (SRG):}  is a Russian-German astrophysical mission designed to carry out an all-sky survey over the soft to medium energy X-ray range. Launched in 2019, on a 7-year mission, SRG carries two instruments—the extended Röntgen Survey with an Imaging Telescope Array (eROSITA) and the Astronomical Röntgen Telescope (ART-XC). Each features electroformed-nickel-replicated X-ray optics.

\begin{figure} [ht]
   \begin{center}
   \begin{tabular}{l} %% tabular useful for creating an array of images 
   \graphicspath{ {./images/} }
   \includegraphics[width=4.0cm]{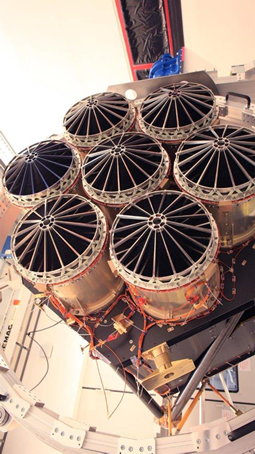}
   \end{tabular}
   \end{center}
   \caption[erosita] 
%>>>> use \label inside caption to get Fig. number with \ref{}
   { \label{fig:erosita} The mirror modules of eROSITA}
   \end{figure}

\emph{eROSITA} is a low-energy instrument ($<10$ keV) with seven mirror modules (see Figure \ref{fig:erosita}), each with 54 gold-coated nickel mirror shells ranging in diameter from 80 mm to 356 mm. The mirror shells are significantly thinner than those for XMM-Newton: only 0.2 mm for the inner shells and up to 0.6 mm for the outer, yet very good imaging performance was achieved, with an average HEW of 16 arcsec at 8 keV over all of the flight modules. All seven mirror modules are co-aligned, giving a total on-axis effective area of over 2000 cm$^2$ at 1 keV. The system focal length is 1.6 m\cite{predehl2016erosita}. 

\emph{ART-XC} is a medium energy instrument (5-30 keV), which like eROSITA consists of seven co-aligned mirror modules each containing mirror shells fabricated via the electroformed nickel replication process\cite{pavlinsky2021art-xc}. The shells in this case are fabricated from a nickel/cobalt alloy that is stronger than pure nickel. They range in size from $\sim$50 mm to 150 mm diameter and in thickness from 0.25 to 0.33 mm, and are coated on the inside with $\sim$10 nm of iridium. There are 28 concentrically nested shells in each module and their focal length is 2.7 m.
The effective area for ART-XC is approximately 65 cm$^2$ per module at 8 keV for an on-axis effective instrument area of 455 cm$^2$. The angular resolution of the ART-XC optics modules is approximately 25 arcsec HEW on axis at 8 keV\cite{gubarev2014calibration}.  However, because SRG is a scanning mission for the first 4 years, it is advantageous to improve the angular resolution off axis, where sources spend most of their time. This can be achieved by defocusing slightly the telescope, degrading the resolution on axis while improving it further out. For ART, the optics are defocused by 7 mm along the optical axis which degrades the on-axis resolution to approximately 30 arcsec HPD.

\subsection{2020s} \label{2020s}
\textbf{IXPE:}  The Imaging X-ray Polarimetry Explorer (IXPE) is a NASA small explorer mission dedicated to x-ray polarimetry\cite{ramsey2021imaging}. Scheduled for launch in December 2021, IXPE features 3 identical telescopes each comprised of a Mirror Module Assembly (MMA) with a polarization-sensitive imaging X-ray detector at its focus.  The IXPE mirror shells are fabricated using the electroformed-nickel-replication process, although in this case the mirror shells are electroformed from a nickel-cobalt alloy which has a higher strength than the conventional pure nickel\cite{ramsey2019ixpe}. Each IXPE mirror module is comprised of 24 concentrically nested mirror shells ranging in intersection diameter from 162 mm to 272 mm. The shells are uncoated, the bare nickel/cobalt alloy giving optimum reflection when integrated over the 2-8 keV IXPE band. Full specifications of the IXPE mirror system is given in Table \ref{tab:ixpe}.

\begin{table}[]
    \centering
    \caption{IXPE Mirror Module Assembly final specification}
    \label{tab:ixpe}
    \begin{tabular}{l|l}
    Parameter & Value \\
    \hline
    \hline
    Number of mirror modules & 3 \\
    Number of shells per module & 24 \\
    Focal length  & 4 m\\
    Total shell length & 600 mm \\
    Range of shell diameters & 162-272 mm\\
    Range of shell thicknesses & 0.18-0.25 mm \\
    Shell material & Electroformed nickel-cobalt alloy \\
    Effective area per module &  166 cm$^2$ (at 2.3 keV); $>175$ cm$^2$ (3-6 keV)\\
    Angular resolution (HEW) & $\leq 28$ arcsec \\
    Module Mass & 32 kg \\
    \end{tabular}
\end{table}

To keep within the weight budget of the small explorer program while maximizing effective area the mirror shells are quite thin and this posed a challenge for the assembly process as it is quite easy to distort the shells during integration and bonding. The final design for the IXPE mirror modules called for the shells to be supported one end only by a rigid spider that would carry the loads to the support structure on the optical bench. A second spider on the rear of the module featured combs that were not bonded to the shells but would limit shell excursions during launch.

\begin{figure} [ht]
   \begin{center}
   \begin{tabular}{l} %% tabular useful for creating an array of images 
   \graphicspath{ {./images/} }
   \includegraphics[width=7.0cm, angle=270,origin=c]{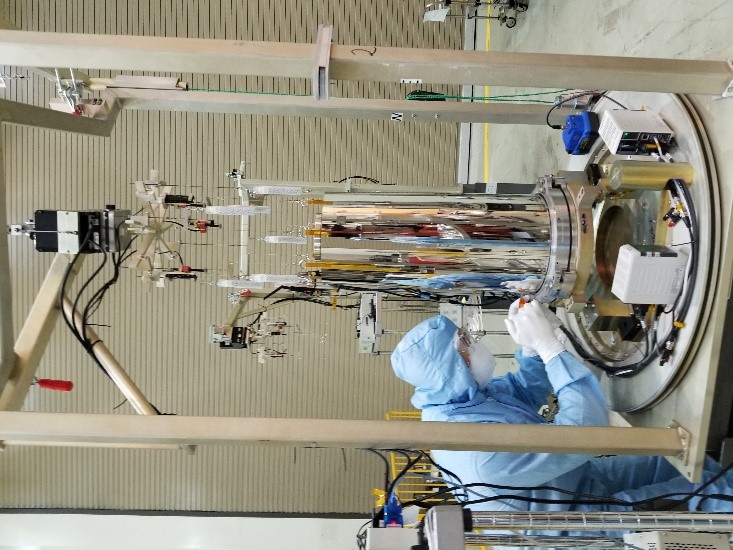}
   \end{tabular}
   \end{center}
   \caption[ixpe] 
%>>>> use \label inside caption to get Fig. number with \ref{}
   { \label{fig:ixpe} An IXPE mirror module assembly station}
   \end{figure}

For shell integration, an approach used for both XMM/Newton and eROSITA/SRG was adapted. In this, the individual mirror shells were suspended from 9 support wires, 6 serving as mass off-loaders and 3 adjustable via piezo electric actuators. With the shells floating in grooves of combs bonded to the front support spider the 3 actuators were adjusted under computer control to optimize the centering and alignment of each mirror shell as measured by a set of proximity sensors that monitor radial displacements of the outside of the shell. When the performance is optimized a low-shrinkage epoxy is injected into the comb slots and the assembly is left to cure overnight\cite{bongiorno2021assembly}. Figure \ref{fig:ixpe} shows an outer shell being bonded on the IXPE assembly station, one of four stations built for the IXPE program to assemble three flight and one spare mirror module in parallel.

When completed the three flight mirror modules were calibrated in the Marshall Space Flight Center 100-m test facility, giving the effective area and angular resolution reported in Table \ref{tab:ixpe}. Factoring in the spatial resolution of the gas-filled focal plane detectors and spacecraft aspect uncertainty, the final system-level angular resolution for IXPE was 28 arcsec HEW, averaged over the 3 mirror modules, meeting requirements. The IXPE mission will be a pathfinder for the new field of X-ray polarimetry, and in its 2-year baseline mission will look at representative examples of many classes of X-ray sources in year 1 with detailed follow-on studies in year 2.

\textbf{XRISM:}  The X-Ray Imaging and Spectroscopy Mission (XRISM) is a follow-on the to the Hitomi mission that ended prematurely. Due to launch in 2022, XRISM will feature two telescopes -  one a wide field imager with an array of CCDs and the other a narrow-field imager with an array of microcalorimeters that will deliver imaging spectroscopy over a 0.3-12 keV energy band with a resolution of less than 7 eV at 6 keV. The X-ray mirror assemblies for XRISM are essentially direct copies of those used in Hitomi, but with slight modifications in the foil replication process for the larger diameter reflectors to improve angular resolution\cite{okajima2020development}. Each of the two identical mirror modules will contain 203 nested shells made from heat-formed aluminum foils of thickness varying from 150 µm to 300 µm. To get the desired surface roughness the formed foils go through an epoxy replication process (10 $\mu$m of epoxy) off gold-sputtered Pyrex mandrels\cite{soong2014astro-h}. The XRISM mirror modules have a focal length of 5.6 m and a measured per-module effective area of nearly 600 cm$^2$ at 1.5 keV and 370 cm$^2$ at 8 keV.

\textbf{eXTP:} The enhanced X-ray Timing and Polarimetry mission (eXTP), is a flagship X-ray astronomy mission led by the Chinese Academy of Sciences with a substantial European contribution. Scheduled for launch in 2027, and currently in phase B, eXTP will have a suite of 4 instruments devoted to imaging, timing spectroscopy and polarimetry. Two of these instruments, the Spectroscopic Focusing Array (SFA) and the Polarimetry Focusing Array (PFA), feature grazing incidence optics. The optics for the SFA and the PFA are very similar to each other in design\cite{basso2019eXTP}, with response from 0.5-10 keV, focal lengths of 5.25 m, and outer diameter constrained by a mirror module envelope of around 600 mm. The principal difference is in the requirement for angular resolution which is 1 arcminute for the SFA and 30 arcsec (with a goal of 15 arcsec) for the PFA.  The baseline fabrication technique for all the optics is electroformed nickel replication, and the only planned difference between the SFA and PFA modules would be in the difference in shell thickness -  a factor of two greater for the PFA to provide a stiffer shell for improved angular resolution.  At the current stage of the design, there are expected to be around 40 mirror shells per mirror module, with the shell thickness ranging from 0.12 mm to 0.22 mm for the SFA and 0.23 mm – 0.44 mm for the PFA optics. The effective area per module is expected to be greater than 820 cm$^2$ at 2 keV and greater than 550 cm$^2$ at 6 keV. The final eXTP payload will contain 9 mirror modules in the SFA and 4 mirror modules in the PFA.

\textbf{ATHENA:}  As has been the case for every major scientific mission in the X-ray band the definition of the ATHENA mission started very early. Even before the launch of XMM-Newton a very ambitious telescope concept was studied by the first telescope working group. This originated at a meeting in 1996 in Leicester UK, where a report named ‘Next Generation X-ray Observatories’ was formulated.  This telescope had a focal length of 50 m and provided up to 30 square meters of effective area at 1 keV and a sensitivity 200 times better than XMM-Newton. The scientific aim of the mission was to study the hot and energetic universe using high and medium resolution spectroscopy with a wide field of view and good spatial resolution and unprecedented effective area. These design studies eventually became the foundation for the XEUS mission concept which was presented to ESA as the candidate for the second L mission in the new cosmic visions program targeting missions in the 2015-2025 time frame. In the same period NASA studied the Constellation-X mission concept. In 2008 the two concepts were merged to form a joint ESA/NASA/JAXA mission IXO – The International X-ray Observatory, but this was abandoned in 2011 and ESA continued with a cost reduced modification which became the ATHENA mission concept and was eventually selected for The L2 mission in 2014. 

\begin{figure} [ht]
   \begin{center}
   \begin{tabular}{l} %% tabular useful for creating an array of images 
   \graphicspath{ {./images/} }
   \includegraphics[width=7.0cm]{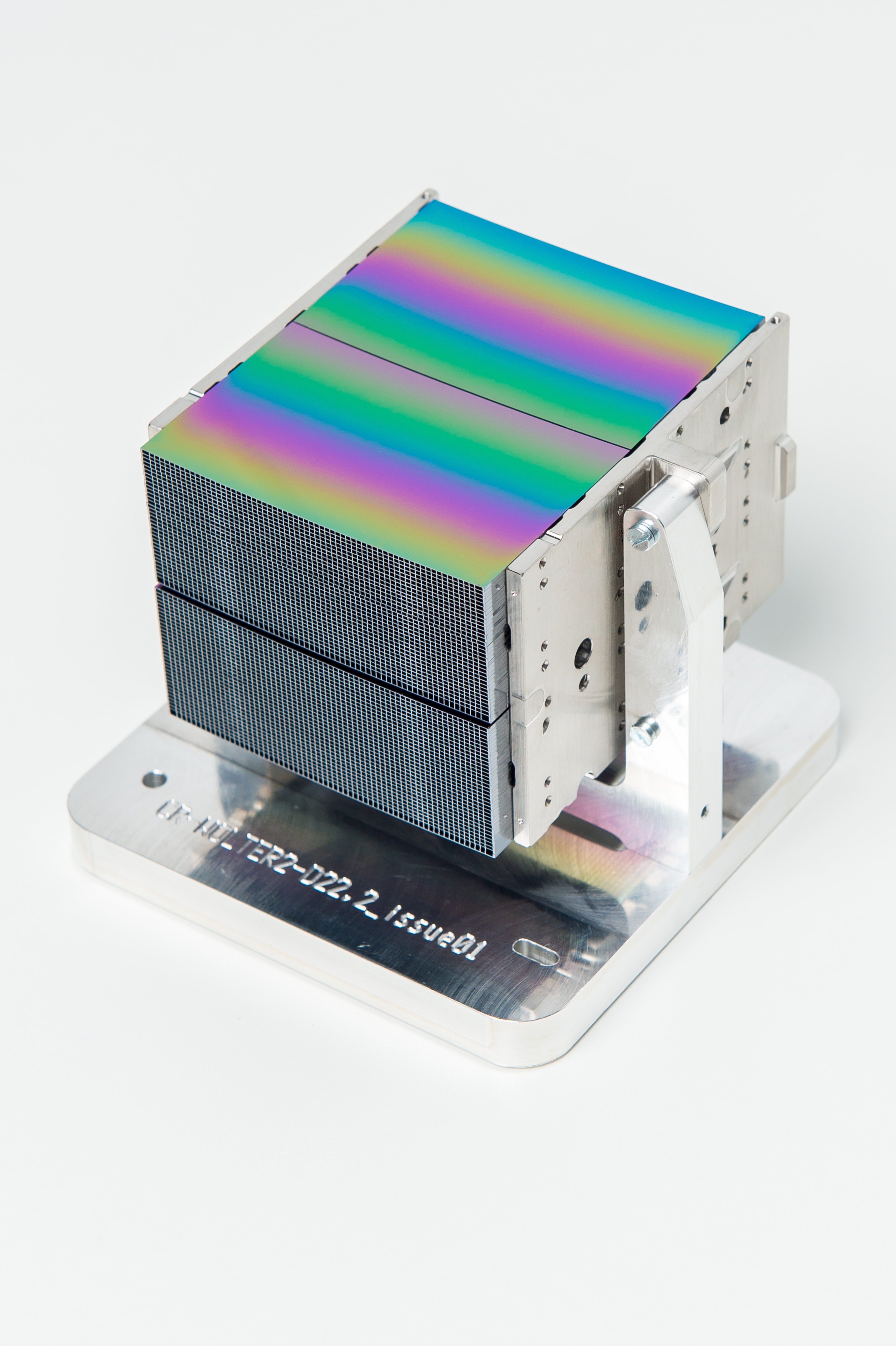}
   \end{tabular}
   \end{center}
   \caption[athena] 
%>>>> use \label inside caption to get Fig. number with \ref{}
   { \label{fig:athena} A silicon pore module}
   \end{figure}

The ATHENA telescope design has undergone several iterations but is now settled at a single telescope 2.5 m in diameter with 12 m focal length, aiming at 1.4 m$^2$ effective area at 1 keV and 0.25 m$^2$ at 6 keV\cite{bavdaz2021athena}. The scientific aim remains the same and very much builds on the heritage of XMM-Newton and Chandra. Since the very beginning it was clear that a novel telescope concept was required to meet the stringent mass restrictions while achieving the high effective area and a required imaging resolution of 5 arcsec. The technology developed and selected for this was the Silicon Pore optics making use of the very smooth and flat Silicon wafers produced in the semiconductor industry\cite{collon2021silicon}. In this implementation, the silicon wafers are etched to provide support ribs and a stack of wafers are cold-welded together under pressure to form a block of individual reflecting channels or pores (see Figure \ref{fig:athena}). The net result is an optic that is extremely rigid, yet very light weight. This technology has been under development for more than a decade by ESA. The mission requires on order 90000 individual mirror elements robotically stacked into 600 units coaligned to a common focus. The resulting telescope parameters are given below (Table \ref{tab:athena}). It is expected that a Mission Adoption Review will take place in 2023 and the launch is foreseen in 2030´s.

\begin{table}[]
    \centering
    \caption{Parameters of the Athena Mirror System}
    \label{tab:athena}
    \begin{tabular}{c|c}
    Parameter & Value \\
    \hline
    \hline
    Focal Length    &  12 m \\
    Outer Diameter & 2.6 m \\
    Graze Angles & 21 - 89 arcmin \\
    Effective Area & 1.7 m$^2$ at 1 keV \\
    Number of silicon pore modules & 600 \\
    Angular Resolution (HEW) & 5 arcsec \\ 
    \end{tabular}
\end{table}

\subsection{Future} \label{future}
\textbf{Lynx:}  Looking beyond currently-funded missions to a future follow-on mission to Chandra reveals significant technical challenges. The Lynx X-ray observatory\cite{gaskin2019lynx} is one of four strategic mission concepts that were studied for inclusion in NASA’s 2020 decadal survey. Scientific considerations levelled requirements of Chandra-level angular resolution, to avoid source confusion and facilitate counterpart identification, an effective area 20 x larger than Chandra, and significantly-improved off axis response. Achieving this, while being able to launch the observatory on a standard heavy-lift rocket calls for significant changes to mirror technology. Specifically, much lighter (thinner) reflectors must be used to limit mass and to permit the dense packing necessary to achieve the required effective area. However, thinner reflectors have much less rigidity and this makes it much harder to achieve high angular resolution when internal substrate stresses and mounting forces (such as from epoxy shrinkage) are considered.

Various technologies are being considered for the 3-m-diameter Lynx mirror system. Among these are conventional (Chandra-like) full shell optics, directly fabricated, yet from substrates an order of magnitude thinner than those of Chandra. A second technology under investigation is that of active segmented optics where $\sim$0.5-mm-thick glass substrates are coated with an array of piezoelectric surface-tangential actuators which can be used to modify the figure of the reflector after mounting\cite{cotroneo2018progress}.  A third technology, that of silicon meta-shell optics\cite{zhang2018astronomical}, uses monocrystalline silicon, an essentially stress-free material that is ground, lapped and polished and then sliced to give a very thin substrate of dimensions 100 mm x 100 mm x 0.5 mm. After ion beam figuring, the mirror segment is coated and then point bonded (4 locations) to a silicon support structure. This is segmented approach lends itself to mass production for the $\sim$37.5k reflectors that would be needed for Lynx. The meta-shell optic approach has been baselined for study of a Lynx design reference mission.

A Lynx technology roadmap shows that if selected in 2024, the mission could go through critical design review in $\sim$2030 and launch in the mid 2030’s\cite{gaskin2019lynx}.

%\end{document}

\section{Conclusion}
%\begin{document}
The use of grazing-incidence X-ray optics has transformed the field of X-ray astronomy, taking us from the few hundred sources in the first X-ray-sky catalog in 1972 to the equivalent of more than 10,000 sources per square degree in the current Chandra deep fields. Getting there has posed numerous technological challenges, all of which were successfully overcome. The next generation of instruments will be equally (or possibly even more) challenging as ever more collecting area is demanded along with high angular resolution -  and all within manageable budgets and launch-able masses. The future of grazing-incidence X-ray optics promises to be just as exciting as the past $\sim$6 decades have been.

%\end{document}

\bibliographystyle{unsrt}
\bibliography{bibliography}
\end{document}